\documentclass{aa}  
\usepackage[switch]{lineno} %\linenumbers

\usepackage{graphicx}
\usepackage{subcaption}
\usepackage{color}

\usepackage{txfonts}
\usepackage[colorlinks=true,linkcolor=blue,filecolor=blue,citecolor=blue,urlcolor=blue]{hyperref}
\usepackage{xspace}

\usepackage{txfonts}
\usepackage{longtable}
\usepackage{amsmath}
\usepackage{geometry}
\usepackage{changepage}
\usepackage{array}
\usepackage{tabularx}
\usepackage[utf8]{inputenc}
\usepackage{float}
\usepackage{caption}
\usepackage{subcaption} % 用于子图
\usepackage{booktabs}
\usepackage{listings}
\usepackage{times}
\usepackage{natbib}
\usepackage{amssymb}
\usepackage{bm}
\usepackage{newtxtext}
\usepackage[T1]{fontenc}
\usepackage{ae,aecompl}
\usepackage{threeparttable}
\usepackage{amsmath}
\usepackage{ulem}
\usepackage{txfonts}

\usepackage{amssymb}	% Extra maths symbols
\usepackage{url}

\def\apjl{{ApJ}}                % Astrophysical Journal, Letters
\def\apjs{{ApJS}}               % Astrophysical Journal, Supplement
\def\ao{{Appl.~Opt.}}           % Applied Optics
\def\aap{{A\&A}}                % Astronomy and Astrophysics
\def\kms  {km~s$^{-1}$}

\def\kmsperkpc  {km~s$^{-1}$~kpc$^{-1}$}

\bibpunct{(}{)}{;}{a}{}{,}
\usepackage{hyperref}
\usepackage{longtable}
\usepackage{booktabs}
\definecolor{dkgreen}{rgb}{0,0.6,0}
\definecolor{gray}{rgb}{0.5,0.5,0.5}
\definecolor{mauve}{rgb}{0.58,0,0.82}
\definecolor{golden}{rgb}{0.86,0.65,0.01}

\begin{document} 

\titlerunning{Galactic OC warp and precession }
\authorrunning{Peng $\&$ He}

   \title{A Slowly Flattening Milky Way Stellar Disk:\\ Investigating Galactic Warping through Dynamical Orbital Inclinations of Open Clusters}

   %\subtitle{I. Overviewing the $\kappa$-mechanism}

   \author{Liming Peng\inst{1} \and Zhihong He\inst{1}\fnmsep\thanks{E-mail: hezh@mail.ustc.edu.cn}}

   \institute{School of Physics and Astronomy, China West Normal University, No.1 Shida Road, Nanchong 637002, People's Republic of China}
\date{Received  / Accepted }

  \abstract
{By evaluating angular momentum directions of open cluster (OC) samples across various Galactocentric radii, we assessed their orbital plane inclinations. Our findings reveal that, without considering the local tilt of the Galactic disk near the sun, our results are consistent with previous studies on Classical Cepheids (CCs). Notably, the warp precession derived from OCs closely mirror those of CCs. Nonetheless, we observed a systematic deviation between the geometric and dynamic warps, attributable to the tilt of the local disk. We identified a systematic vertical motion in the local region, associated with the warping feature near the solar vicinity. Ignoring this motion leads to underestimates of orbital plane inclinations compared to those derived from geometric positions. Our study indicates consistency between the inclinations derived from orbital dynamics and geometric positions at a vertical velocity of the sun relative to the Galactic mid-plane of $V_{z\odot}$ =~9.4~$\pm$~0.2~\kms. This value is approximately  2~\kms higher than the historically estimated solar peculiar motion, $W_{\odot}$, primarily due to a $\theta_i$~$\approx$~0.6$^{\circ}$ tilt of the local plane. Analysis suggests that previous estimates of the Galactic disk's warping precession rate may have been overestimated due to local warping influences. The findings indicate that the precession oscillates around zero and that the Galactic warp is progressively flattening. Additionally, the line of nodes tends to become consistent across various Galactocentric radii over a timescale of 100-200 million years.}


\maketitle{}
%
%-------------------------------------------------------------------
\section{Introduction}\label{sec:intro}
Galactic warping primarily manifests as distortions or bending at the edges of a galaxy's disk, and most evident in spiral galaxies~\citep[e.g.][]{Bosma81,Ruiz02}. The causes of such warping remain partially understood, with several theories proposed. One leading hypothesis suggests that gravitational interactions with neighboring galaxies can distort galactic structures~\citep[e.g.][]{Shang98,Weinberg06,Schonrich18}. Another involves the asymmetric distribution of dark matter halos~\citep[e.g.][]{Sparke88,Nelson95,Han23}, whose mass substantially surpasses that of visible matter, leading to disk warping. Internal dynamic processes within the galaxy disk, such as the collapse of gas clouds and formation of new stars can disrupt galactic equilibrium~\citep{Ceverino12}. Additionally, the interaction between galactic rotation and external environments like hot gas halo is implicated in inducing warping~\citep{Roskar10}. These factors collectively highlight the complex interplay of internal and external forces shaping galactic structures.

For extragalactic galaxies, only the warping structures located on the outer regions of edge-on galaxies are observable, with the Milky Way being the only galaxy that allows for an exploration of the tilt of the entire galactic disk. The warping of the Milky Way disk was first discovered in the 1950s~\citep{Burke57,Kerr57}, but it is only in recent years that advancements in surveys such as variable star census~\citep{Chen18,Udalski18} and astrometric surveys~\citep{gaiadr2,gaiaedr3,gaiadr3} have gradually enhanced our understanding of the three-dimensional structure of the disk. 
Currently, the warping and its precession are tracked using three primary observational methods: geometric, kinematic, and dynamical. Geometric methods involve estimating the warp's amplitude by mapping the spatial positions of specific tracers within the Milky Way's disk~\citep[e.g.][]{Chen19,Skowron19,Lemasle22,He23}. Kinematic methods assess warping and precession by analyzing perturbations in the vertical velocities of stars in the Galactic plane~\citep[e.g.][]{Poggio18,Poggio20,Cheng20, Zhou24}. Dynamical methods, on the other hand, derive insights from the mean direction of stellar angular momentum, which correlates with the inclinations and orientations of stellar orbits~\citep[][]{Dehnen23}. These multifaceted approaches have greatly enhanced our comprehension of Galactic warps.

However, due to age discrepancies among tracers, incomplete samples caused by extinction, and uncertainties in solar motion velocities, there are significant discrepancies in the study of the inclination angle of the Milky Way disk warping with respect to the Galactocentric radius. Currently, most studies utilizing geometric and kinematic approaches place the origin of the warping beyond the solar radius, ranging from 8 to 12~kpc ~\citep[e.g.][]{Chen19,Skowron19,Romero19,Cheng20,Chrobakova20}. Among them, ~\citet{Romero19} used OB and AGB stars as samples and found differences in warping origins between young and old stellar populations, initiating at ~12-13~kpc and ~10-11~kpc for young and old stars, respectively. However, some studies indicate the warping may start in the inner disk~\citep{Ralph18,Skowron19b,He23}. In our previous study~\citep[][hereafter H23]{He23}, we utilized open star clusters to investigate the geometric and kinematic characteristics of the Milky Way disk from 4 to 14~kpc, revealing varying degrees of warping in both the inner and outer disk. The tilt angle of the disk near the sun was found to be approximately 0.6$^{\circ}$.

Recently, ~\citet{Dehnen23} (hereafter D23) analyzed the distribution of the orbital angular momentum of classical Cepheids (CCs) at different Galactocentric radii, revealing the tilted state of the Milky Way disk and, for the first time, indicating a decreasing trend in precession in the outer regions of the disk. This trend also aligns with the precession pattern identified in H23. In this work, using the OC samples, we reproduced disk precession values consistent with the results from CC sample by applying the same vertical velocity of the sun (Section~\ref{sec:3}). As results obtained from dynamical orbits, the direction of angular momentum is unaffected by its physical distribution, allowing for warping parameters independent of geometric features. However, their research found that the inclination angle derived from dynamical methods does not align with the physical distribution of CCs. 
Interestingly, the differences between the two are systematic: as illustrated in Figure~3 of D23, Cepheid samples exhibit larger (0.5$^{\circ}$ to 1$^{\circ}$) inclinations in their physical distribution compared to those derived from dynamical orbits at Galactocentric distances of 7 to 16~kpc.
%While interstellar extinction and the resulting homogeneous sample distribution can lead to differences between geometric and dynamical results. ~\textbf{However, the impact of extinction on the observations of objects on the northern and southern sides of the Galactic plane is randomized. Consequently, the discrepancy between the observed geometric inclination and the dynamical inclination of the Galactic disk is random rather than systematic.}
This particular issue has not undergone a more thorough investigation, what could be the underlying reason for this systematic discrepancy?

According to the analysis in this study, the primary influence when studying the warping inclination through methods such as tracing the vertical velocity and angular momentum of tracers is the sun's vertical motion relative to the Galactic mid-plane (Section~\ref{sec:4}). When calculating the three-dimensional velocities of Galactic objects, it is necessary to correct for the sun's peculiar motion relative to the local standard of rest (LSR). However, if the orbit in the vicinity of the solar system is not parallel to the Galactic mid-plane (and has an inclination angle $\theta_i$), the tilted orbit introduces an additional vertical systematic velocity to the LSR. This systematic velocity cannot be subtracted by statistically averaging the motions of solar nearby objects. We have discovered that even a small tilt angle of $\theta_i$~=~0.6$^{\circ}$ could introduces a velocity of $V_{\odot} sin\theta_i$~$\sim$~2.5~\kms (not consider precession) in the LSR frame. This value already represents around one-quarter to one-third of the existing solar peculiar motion in $W_{\odot}$ and may leads to systematic biases in warping dynamical quantities relative to the physical characteristics.

To address this issue, we will proceed to validate this analysis using open clusters (OCs). Compared to CCs, the sample size within 5~kpc of the solar system is twice as large for OCs, and those stellar aggregates provide more precise kinematic parameters compared to individual stellar tracers. Previously, H23 obtained the physical distribution and kinematic features of the warping through OCs. Inspired by D23, we will further explore the warping structure from an angular momentum perspective: by averaging the instantaneous angular momentum positions of OCs at different Galactocentric radii, we can infer the tilt of dynamical orbits and compare these results with geometric predictions. 
Considering the factors such as the uncertainties in peculiar motion, the influence of local warping on the LSR due to the sun's local orbital tilt causing vertical systematic velocities, and possible perturbations such as inclination variation and oscillations near the sun, we aim to determine the vertical velocity $V_{z\odot}$ of the solar system relative to the Galactic mid-plane.

The structure of this paper is organized as follows: 
Section~\ref{sec:2} introduces the selected star cluster sample and sketch map of the dynamical inclination. Section~\ref{sec:3} present the warp features based on CC and OC samples when ignore the effect of the local warp.
In Section~\ref{sec:4}, we constrain $V_{z\odot}$ using the dynamical and geometric tilt of the warp, and investigate the inclination and line-of-node (LON) positions of warping under different OC age ranges. 
Section~\ref{sec:5} presents our conclusions and discussions. 
In Appendix~\ref{sec_sketch_map} and ~\ref{sec:appendix_a} we introduce the sketch map of dynamic warping and the statistics of solar peculiar motions, respectively.

\section{Sample and Method}\label{sec:2}
\begin{figure*}
\centering
\includegraphics[width=0.495\linewidth]{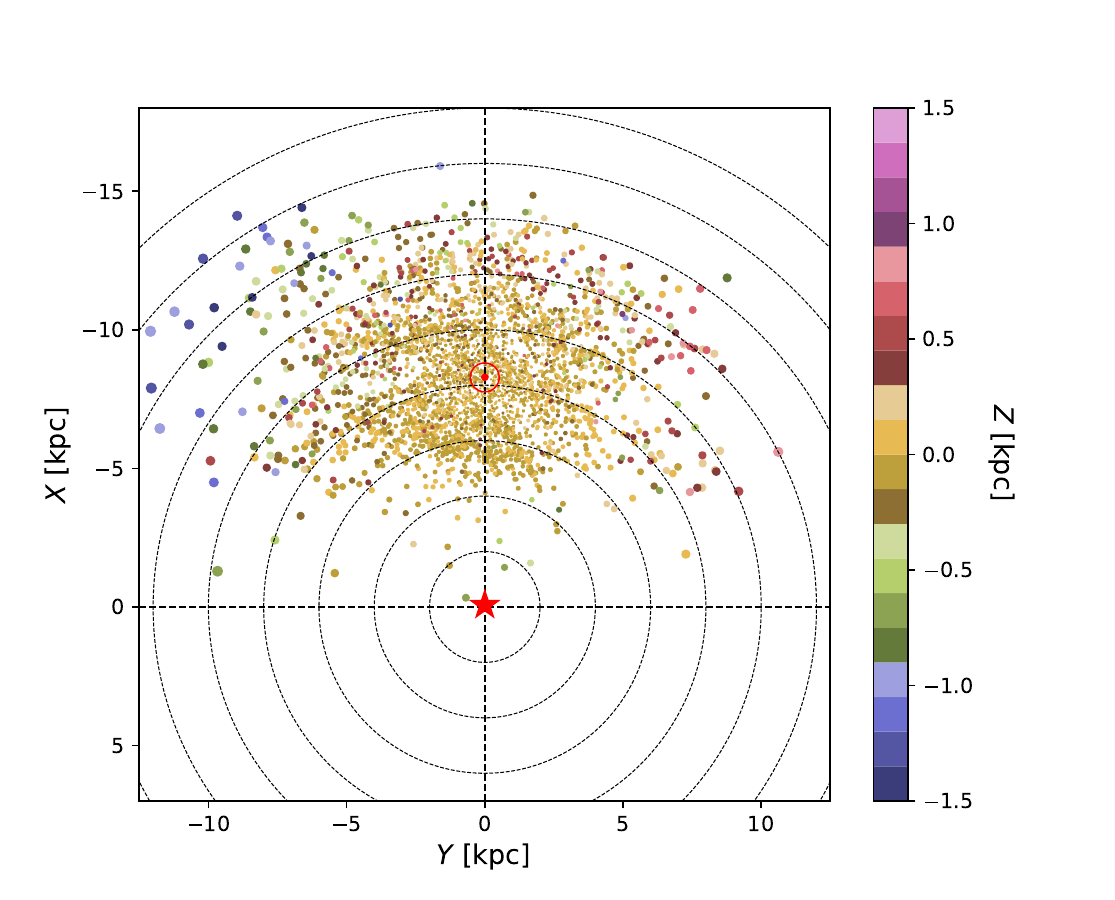}
\caption{Distribution of OC samples in the $X$-$Y$ plane, color-coded by the vertical height. To emphasize the height variation across the plane, the scatter point sizes are scaled proportionally to the square of the distance. The position of the sun is at (- 8.15, 0) kpc, according to \citet{Reid19}.
}
\label{fig1}
\end{figure*}

%\section{Inferred $V_{z\odot}$ under an inclined plane}\label{sec:2}
In our study, OCs were utilized as tracers to map the warping structure, as derived from the comprehensive dataset aggregated by H23 and the works referenced therein, predominantly those of ~\citet{cg20,He23a,He23b}. For the purpose of accurately estimating the angular momentum, our inclusion criteria were meticulously defined: only OCs possessing a vertical velocity uncertainty below 1~\kms and a tangential velocity uncertainty under 20~\kms were included. This approach resulted in a robust sample comprising 3991 OCs, providing a solid basis for a comprehensive averaged dynamical analysis.
The ages of these clusters were derived from isochrone fitting, include 1408 young clusters (<100 Myr), 1985 middle-aged clusters (0.1 to 1~Gyr), and 598 old clusters ($\geqslant$~1~Gyr). 
The distribution of these cluster samples on the Galactic plane is presented in Figure~\ref{fig1}, with the majority of the samples located within a radius of 14~kpc from the Galactic center.
%%%%%%%%%%%%%%%%%%%%%%%%%%figure%%%%%%%%%%%%%%%%%%%%%%%%%%%%%%%%%
%%%

In order to facilitate the study of warping through them, we establish a Cartesian coordinate system with the Galactic Center as the origin (Figure~\ref{fig2}), where the $\emph{Y}$-axis is aligned with the direction of the circular rotation at the sun's position, the $\emph{X}$-axis points away from the sun, and the $\emph{Z}$-axis is orthogonal to the $\emph{X-Y}$ plane (northern Galactic pole). Additionally, we also utilized Galactocentric cylindrical coordinates $\mathrm{R_{GC}}$, $\phi$, and $\emph{Z}$, where Galactic azimuth $\phi$ =~0$^\circ$ aligns with the $\emph{X}$-axis, extending in the counterclockwise direction.
The sun's coordinates are set at (-8.15, 0, 0.005) kiloparsecs, as referenced from ~\citet{Reid19}, and $\phi_{\odot}$ is 180$^\circ$. Then we follow D23 and analyze the Galactic warp from the perspective of angular momentum. 

As shown in Figure~\ref{fig2}, the instantaneous angular momentum direction of each cluster in the Milky Way can be determined by its position and velocity. The angular momentum $\bm{L}$ can be expressed as the cross product of the position vector $\bm{r} (x, y, z) $ and the velocities $ \bm{v}(v_x, v_y, v_z)$: $ \bm{L} = \bm{r} \times {\bm{v}}$, decompose the vectors \( \bm{r} \) and \( \bm{v} \) into their components along three directions.
Then, the angle between the projection of $\bm{L}$ on the $\emph{X-Y}$ plane and the $\emph{X}$-axis is given by $\phi= \mathrm{arctan(L_y/L_x)}$, while the inclination angle $\theta_\mathrm{i}$ between $\bm{L}$ and the $\emph{Z}$-axis determines the orbital inclination. From Figure~\ref{fig2}, we note that $\mathrm{sin\theta_i= {\sqrt{L_x^2+L_y^2}/ \bm{L}}}$, as  $\theta_\mathrm{i}$ is small value, $\theta_\mathrm{i}$ can be approximated as $\sqrt{{L_x^2}/{{L}^2}+{L_y^2}/{{L}^2}}$. In order to compare with the geometric warp and dynamical warp, we used the Galactocentric radius of OCs instead of guiding radius, and we found that the distributions obtained from the two did not have a significant deviation. 

We then perform statistical analysis on OCs within a certain Galactocentric radius range (bin = 0.5~kpc), employing the bootstrap method to calculate the tilt angle and LON direction at different Galactocentric radii, along with their magnitudes and uncertainties by repeatedly sampling from data within that radial range. Considering that the view of the warping in the northern and southern directions within the Galactic plane is opposite when seen from the anti-Galactic center ($\phi$ = 180$^{\circ}$), we define the negative and positive sign of inclination angle when $L_y$ <~0 (Figure~\ref{fig2}-a) and $L_y$ >~0 (Figure~\ref{fig2}-b), respectively; the LON position $\phi_{\mathrm{LON}}$ is determined by the azimuthal angle of the ascending node of intersection on the Galactic mid-plane.

\section{Galactic warping under a flat Local disk}\label{sec:3}

\begin{figure*}
\begin{center}
\includegraphics[width=0.65\linewidth]{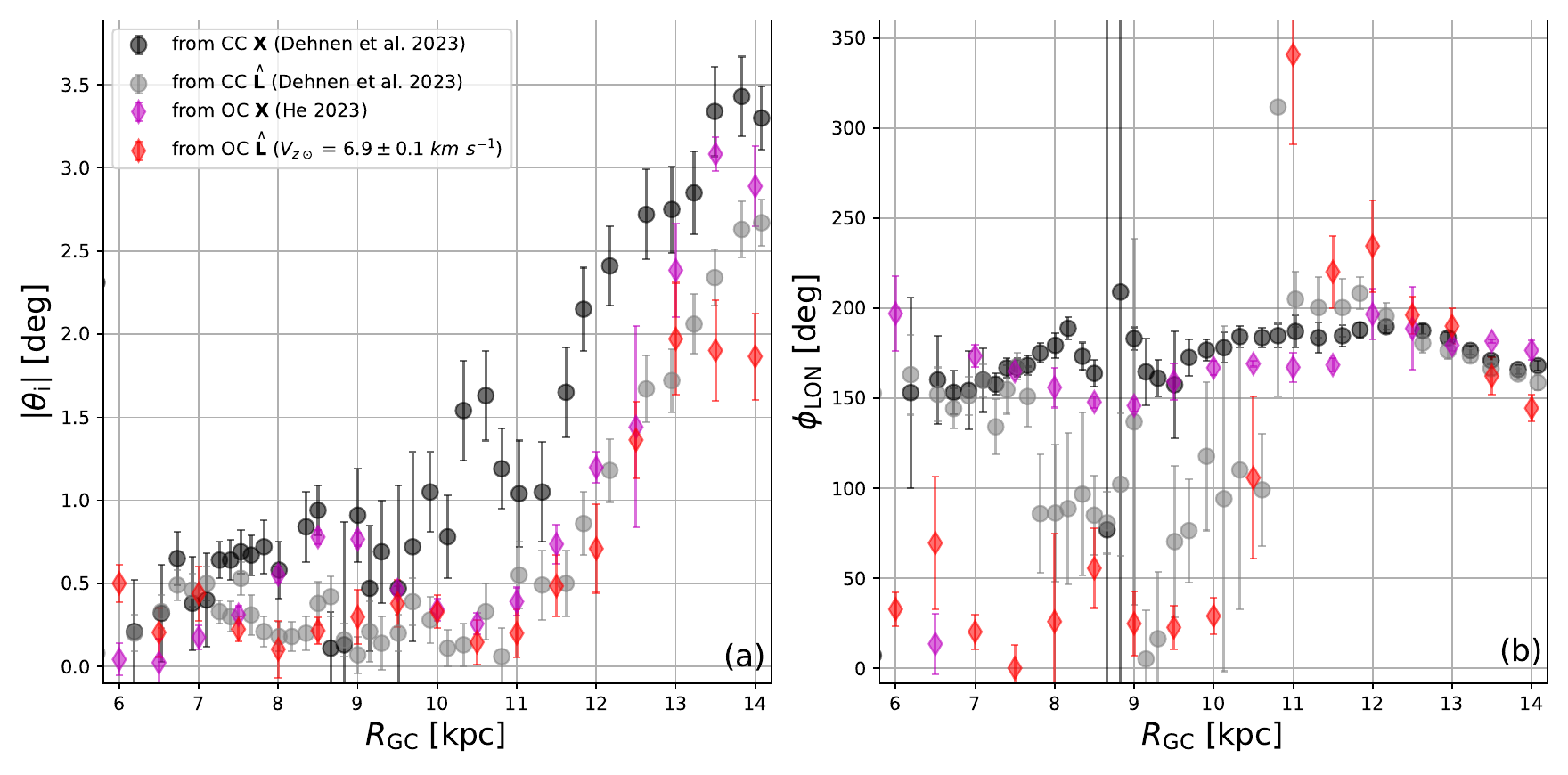}
\caption{Comparison between the OC and CC geometric/dynamical warp, without considering the impact of local warping. Black (CC geometric warp) and grey (CC dynamical warp) points are derived from Figure~3 in D23, and the magenta rhombuses (OC geometric warp) are derived from H23, while red rhombuses are OC dynamical warp. For comparison purposes, $V_{z\odot}$ = 6.9~\kms is adopted same to D23, with an artificial error of 0.1~\kms. Panel (a) illustrates the inclination angle $\theta_\mathrm{i}$ with respect to the Galactocentric radius $\mathrm{R_{GC}}$ for the different tracers and methods. Panel (b) shows the LON position as a function of the $\mathrm{R_{GC}}$.
}
\label{fig3}
\end{center}
\end{figure*}

Based on above method, we then explore the dynamic inclination of the Galactic disk by comparing OC and CC sample data, using the same solar vertical velocity as in D23 for consistency. It is worth mentioning that here we assume that the Galactic plane near the solar system is completely flat ($\theta_i$ = 0), that is to say, $V_{z\odot}$ = $W_{\odot}$ (solar peculiar motion). As illustrated in Figure~\ref{fig3}-a, the warping dynamic inclination angles derived from OC samples align remarkably well with those obtained from CC samples in D23, with both trends displaying nearly identical numerical values and certain radii showing virtually equal inclination values. 
Within 10 to 12.5~kpc, the differences between the geometric tilt measurements for CC and OC are outside the error margins. At other radii, these discrepancies generally fall within acceptable error margins. And the OC geometric tilt is in much better agreement with the dynamic tilt measurement (both OC and CC) than the CC geometric tilt.
However, although the distribution of geometric inclination across various radii in OC samples consistently mirrors the dynamic inclination trend, it is marginally greater than the latter at most radii.

Additionally, the analysis of the LON distribution for geometric warping, as traced by the OC and CC samples, reveals a strong consistency (Figure~\ref{fig3}-b). Both samples exhibit a continuous variation in geometric warping LON within the 150 to 200-degree range. In the outer disk where $\mathrm{R_{GC}} > 12$ kpc, the errors and discrepancies in LON are minimal, underscoring the effectiveness of both sample sets in reliably tracking warping features. Additionally, the trends in the distribution of dynamical warping for OC and CC are generally consistent across most Galactocentric distances. Notable discrepancies are only observed between 6 and 7.5 kpc. In contrast, the outer disk also exhibits exceptional uniformity. However, the dynamical warping LON for both sample sets displays sharp transitions, markedly differing from the continuous LON variation observed in the geometric case. And the difference in the LON between dynamical warping and geometric warping reaches 50 to 150 degrees at certain radii.

\begin{figure*}
\begin{center}
\includegraphics[width=0.675\linewidth]{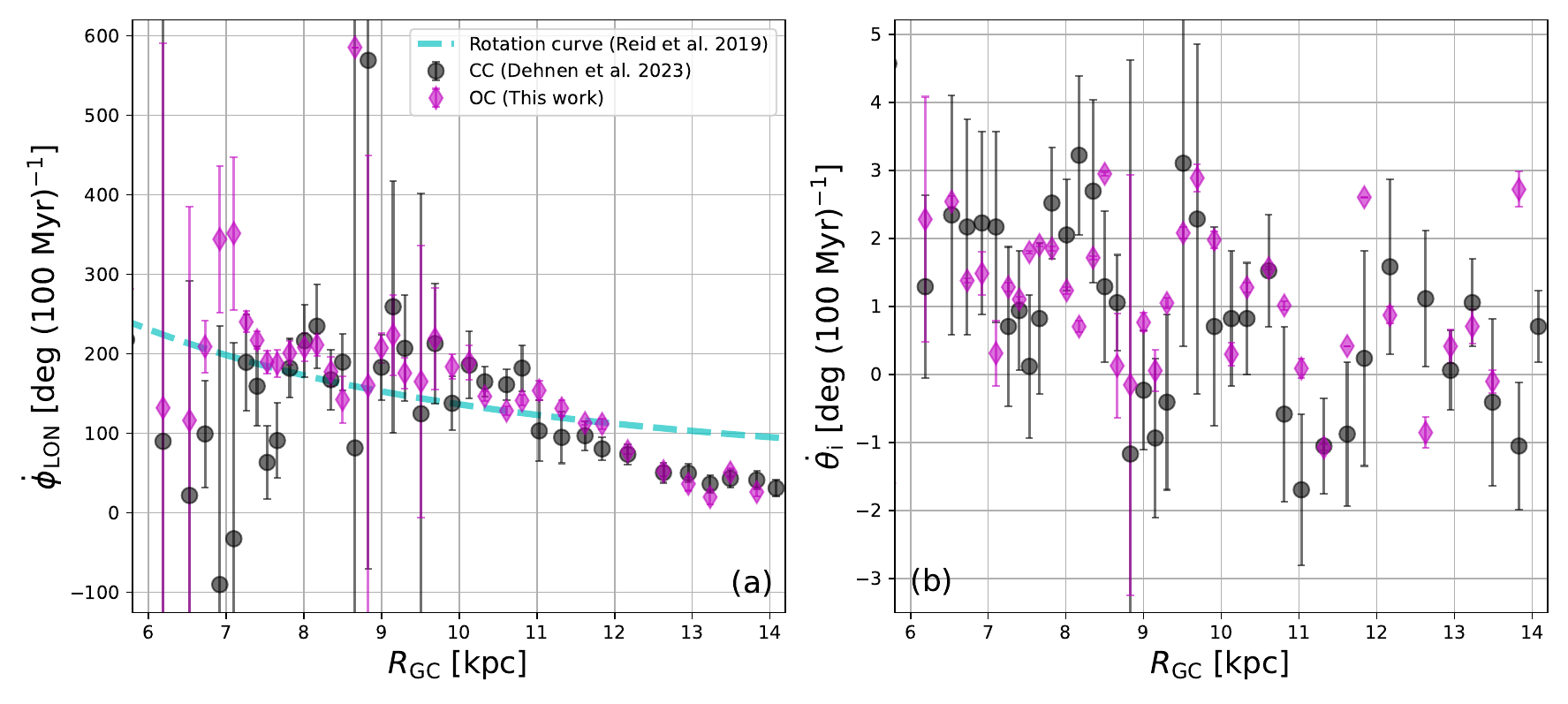}
\caption{Comparison between the OC and CC warp on the precession rate (panel a) and variation rate of the inclination angle (panel b), without considering the impact of local warping. Black points are derived from Figure~3 in D23 (from CC geometric and dynamical warp), and the magenta rhombuses are derived by this work (from OC sample). For comparison purposes, $V_{z\odot}$ = 6.9~\kms is adopted same to D23, with an artificial error of 0.1~\kms; the LON position and inclination angle are also  adopted same to D23.
}
\label{fig4}
\end{center}
\end{figure*}

Dynamical analysis further confirms that the precession rate and inclination variation rate of the Galactic disk are consistent across both tracers~\footnote{In our analysis, here we employed Equation~2 as outlined in H23. The original code is accessible online: \url{https://nadc.china-vo.org/res/r101288/}. Despite H23 and D23 using different samples-open clusters and classical Cepheids respectively-and methodologies, both studies reached consistent results regarding precession, thereby mutually verifying each other's findings. We are also willing to share our modified version of the code developed from this work.}. 
Figure~\ref{fig4}-a demonstrates that the precession rates derived from OC and CC samples closely align. While the inner disk exhibits significant precession error, the central precession rate values for OC and CC fall within 30 $\pm$ 15~\kmsperkpc, equivalent to a change of $\sim$180 degrees per 100~Myr. In the outer disk ($\mathrm{R_{GC}}$ > 12 kpc), both trends highlight a decreasing precession rate with increasing $\mathrm{R_{GC}}$, presenting nearly identical values with some precession values being almost equivalent. As indicated in Figure~\ref{fig4}-b, despite the small magnitude of inclination variation rate, values from OC and CC are consistent, with notably higher values in the inner disk ($\mathrm{R_{GC}}$ < $\sim$10~kpc) compared to the outer disk. 
These consistencies derived using different tracers (OC and CC) and methodologies (D23, H23), underscore robust and consistent warping characteristics of the Galactic disk. These findings contribute significantly to our understanding of Galactic dynamics and structure.

However, two critical aspects warrant careful consideration. Firstly, as outlined in Section~\ref{sec:2}, geometric and dynamical warps may share the same magnitude of $\theta_i$ at given radii, but differences in their LON lead to significant angular discrepancies. Specifically, when LON differences exceed 90 degrees, the actual deviation in theta falls within the range of $|\theta_i|$ < $\Delta {\theta_i}$ < 2 $|\theta_i|$. This results in a disparity between geometric and dynamical inclination angles that is notably greater than the variation depicted in Figure~\ref{fig3}-a, attributable to LON differences. Figure~\ref{fig3}-b further reveals that dynamical LON distributions exhibit discontinuous and complex jumps when local disk tilts are unaccounted for. Secondly, in the absence of local warp considerations, the Galactic disk shows pronounced precession at $\mathrm{R_{GC}}$ < 12~kpc, with values approach and even exceed the rotational speed of the Milky Way, posing challenges to current theoretical models.

﻿
Despite several potential influencing factors-such as extinction, perturbations from satellite galaxies, and Galactic disk precession-these do not sufficiently account for the significant discrepancies observed. OCs and other stellar groups are distributed extensively across the Galactic plane, rendering their geometric distribution vulnerable to extinction effects within the Galactic disk. Despite this susceptibility, extinction minimally affects the dynamic warping of the disk. This results in fluctuations between the geometric and dynamic tilt angles at varying radii. However, these fluctuations do not lead to systematic differences in the overall tilt angle of the disk. Furthermore, ~\citet{Bekki12} has simulated the impact of the Large Magellanic Cloud on the Galactic disk, noting a pole shift effect of approximately 7 $\mu as$/yr. To match the observed differences, an increase by at least one order of magnitude would be necessary. Additionally, current observations indicate that Galactic disk precession is less than 10~\kmsperkpc ~\citep[e.g.][]{Corredoira21}, with the possibility of negative precession~\citep[][]{Huang24}, indicating that geometric inclination does not exceed dynamical inclination markedly.

We also evaluated recent findings by ~\citet{Zhou24} regarding warp precession using CC. They calculated the precession rate of the warp located at 12 to 14~kpc from the Galactic center, reporting a value of 4.9~$\pm$~1.6~\kmsperkpc. By adopting the same solar motion~\citep[][$W_\odot$ = 8.59~$\pm$~0.28~\kms]{gc20210,Drimmel23} used by ~\citet{Zhou24}, we conducted a study on the precession of the warp traced by OCs within the same radial range. Our findings yielded a precession rate of 5.1~$\pm$~3.4~\kmsperkpc. This suggests that the differences in results from various researchers studying warp precession with CC primarily stem from their choices of $W_\odot$ and the radial distance range. Provided that the local warp is not considered and that $W_\odot$ is consistent, our analysis reaffirms the significant consistency between the OC and CC datasets, as well as among the precession calculation methods of D23, H23, and ~\citet{Zhou24}. Therefore, our subsequent analysis will focus on Galactic warps when consider minor tilts in the solar neighborhood disk.

%%%%%%%%%%%%%%%%%%%%%%%%%%figure%%%%%%%%%%%%%%%%%%%%%%%%%%%%%%%%%
%%%%%%%%%%%%%%%%%%%%%%%%%%figure%%%%%%%%%%%%%%%%%%%%%%%%%%%%%%%%%

%%%%%%%%%%%%%%%%%%%%%%%%%%figure%%%%%%%%%%%%%%%%%%%%%%%%%%%%%%%%%
%%%%%%%%%%%%%%%%%%%%%%%%%%figure%%%%%%%%%%%%%%%%%%%%%%%%%%%%%%%%%
%~\textbf{The black points are derived from Figure 3 in D23, showing di/dt and dLON/dt; the magenta points are calculated using OC samples and Function 2 from H23. For comparison purposes, here we used the RGC, LON, and inclination values from the CC samples in D23 (RGC, LONx, and ix), adopting Vz⊙ = 6.9 km/s from D23 with an artificial error of ±0.1 km/s.}

%%%%%%%%%%%%%%%%%%%%%%%%%%figure%%%%%%%%%%%%%%%%%%%%
\section{Local tilt and OC dynamical warping}\label{sec:4}

Traditionally, when solving for $v_x$, $v_y$, and $v_z$, the typical approach involves adding the solar peculiar motion ($U_\odot$, $V_\odot$, $W_\odot$) and the LSR rotation velocity $\Theta_0$. 
However, the precise details of solar peculiar motion have varied across different studies, influenced by methodological differences and potential systematic effects, leading to some variation in the values reported.
As known from Appendix~\ref{sec:appendix_a}, the published values in $W_\odot$ often range from 6 to 9~\kms, while $U_{\odot}$ are range from 5 to 12~\kms, and $V_{\odot}$ + $\Theta_0$ are range from 240 to 270~\kms. 
Since the typical vertical Coordinate of the clusters are small, the averages of $L_x$  and $L_y$  are both dominated by the terms involving $v_z$.
So that the magnitude of $\theta_\mathrm{i}$ strongly depends on the choice of the solar vertical motion, with the radial and tangential velocities having minimal impact, even outside of their nominal range. Simultaneously, the selection of the solar Galactocentric distance (8 to 8.5~kpc) and the Galactic height (3 to 30~pc) has a negligible effect on the calculation of this tilt angle. In our study, more importantly to considering the local warping near the sun ($\theta_i \approx$ 0.6$^{\circ}$), which leads to an upward systematic motion of the LSR (H23,  $\sim$~2~\kms). At the same time, other factors such as inclination variation and/or local perturbations may also influence the systematic motion affecting the vertical velocity of the solar system. By combining the above factors, we defined a vertical velocity of the solar system relative to the mid-plane of the Galactic disk, which denoted as $V_{z\odot}$ = $W_{\odot}$+$V_{z_{LSR}}$, where $V_{z_{LSR}}$ is the vertical systematic motion between LSR and Galactic mid-plane.
%. In our study, the solar vertical velocity $V_{z\odot}$ is set within a certain range (6 to 12 ~\kms) at $V_{z\odot}$ 

\begin{figure*}
\begin{center}
\includegraphics[width=0.705\linewidth]{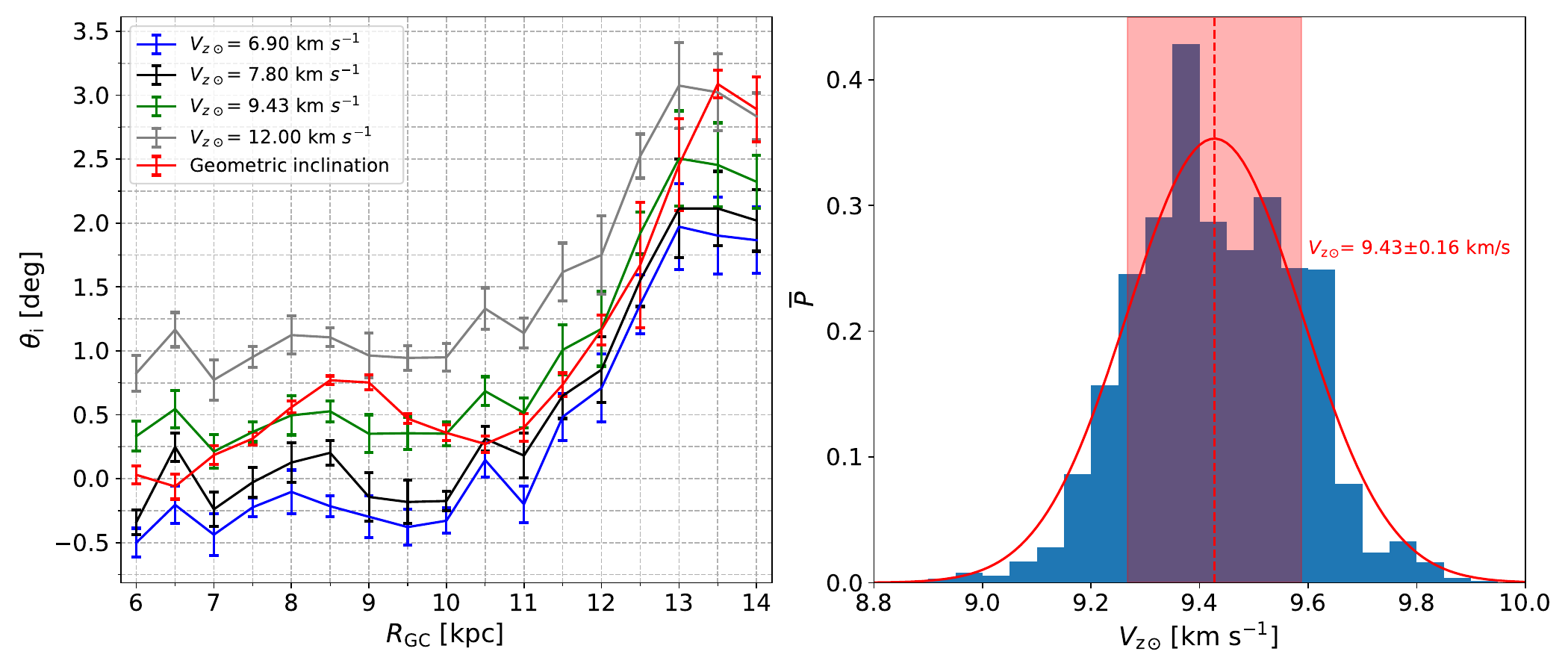}
\caption{Left panel: the variation of the orbital inclination angle $\theta_\mathrm{i}$ of OCs with different $V_{z\odot}$ as a function of the Galactocentric radius. The blue, black, green, and grey lines correspond to $V_{z\odot}$ values of 6.90, 7.80, 9.43, and 12.0~\kms, respectively, while the red line represents the geometrically obtained results from H23 using OCs as samples. The right panel compares the dynamical orbital inclination angle with the geometric inclination angle using a maximum likelihood function to constrain $V_{z\odot}$. The parameter $\bar{P}$ represents the average probability of similarity between the two in each bin (0.05~\kms), resulting in the best gaussian fit (red curve $V_{z\odot}$ =~9.43~$\pm$~0.16~\kms) .
}
\label{fig5}
\end{center}
\end{figure*}

\begin{figure*}
\begin{center}
\includegraphics[width=0.495\linewidth]{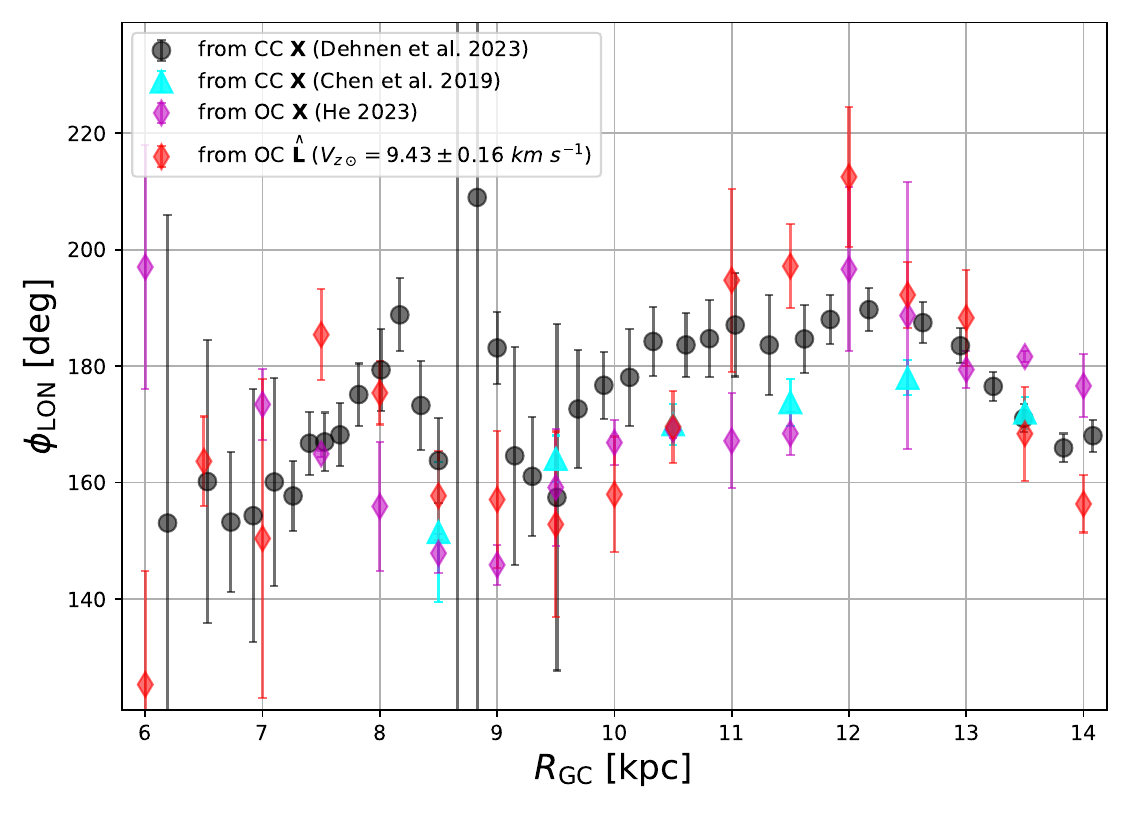}
\caption{Comparison between the OC and CC geometric/dynamical warp on LON positions. Black points are from D23 (CC geometric warp), Cyan triangles are from ~\citep[CC geometric warp,][]{Chen19}, the magenta rhombuses are from H23 (OC geometric warp), while red rhombuses are OC dynamical warp. Different from Figure~\ref{fig4}, $V_{z\odot}$ =~9.43~$\pm$~0.16~\kms is adopted here.
}
\label{fig6}
\end{center}
\end{figure*}

Using the method described above, the results obtained are as shown in Figure~\ref{fig5}: on the left panel, the red line segment represents the results obtained geometrically using OCs as a sample from H23, while the remaining four line segments are the results obtained in this work from the perspective of angular momentum under different $V_{z\odot}$ values. It can be observed that, at the same Galactocentric distance, the tilt of the disk warping progressively increases with increasing $V_{z\odot}$. Additionally, as $V_{z\odot}$ exceeds 7.8 ~\kms, a inclined structure with a positive sign of tilt angle appears at the Galactocentric distance of the solar system; and as $V_{z\odot}$ increases to 12~\kms, the tilt angle of the disk dynamical orbit exceeds the geometric tilt angle, while the inner disk does not exhibit reverse tilting ($L_y$ <~0). Therefore, by identifying a group of dynamical orbits and physical tilt angles closest to each other from multiple curves, the optimal value of $V_{z\odot}$ can be determined. We utilized maximum likelihood estimation to obtain the probability distribution between the two values. Ultimately, as shown in the right panel of Figure~\ref{fig5}, $V_{z\odot}$ = 9.43 $\pm$ 0.16 ~\kms provides the closest match between the two datasets.
In this situation, it is noteworthy that the LON position of the OC dynamic orbit aligns with the LON of the geometric warp of both OC and CC (Figure~\ref{fig6}). This alignment represents a significant improvement compared to the scenario where local warp effects are not considered (Figure~\ref{fig3}-b).

%%%%%%%%%%%%%%%%%%%%%%%%%%figure%%%%%%%%%%%%%%%%%%%%%%%%%%%%%%%%%

%%%%%%%%%%%%%%%%%%%%%%%%%%figure%%%%%%%%%%%%%%%%%%%%%%%%%%%%%%%%%
%%%%%%%%%%%%%%%%%%%%%%%%%%figure%%%%%%%%%%%%%%%%%%%%%%%%%%%%%%%%%

\begin{figure*}
\begin{center}
\includegraphics[width=0.705\linewidth]{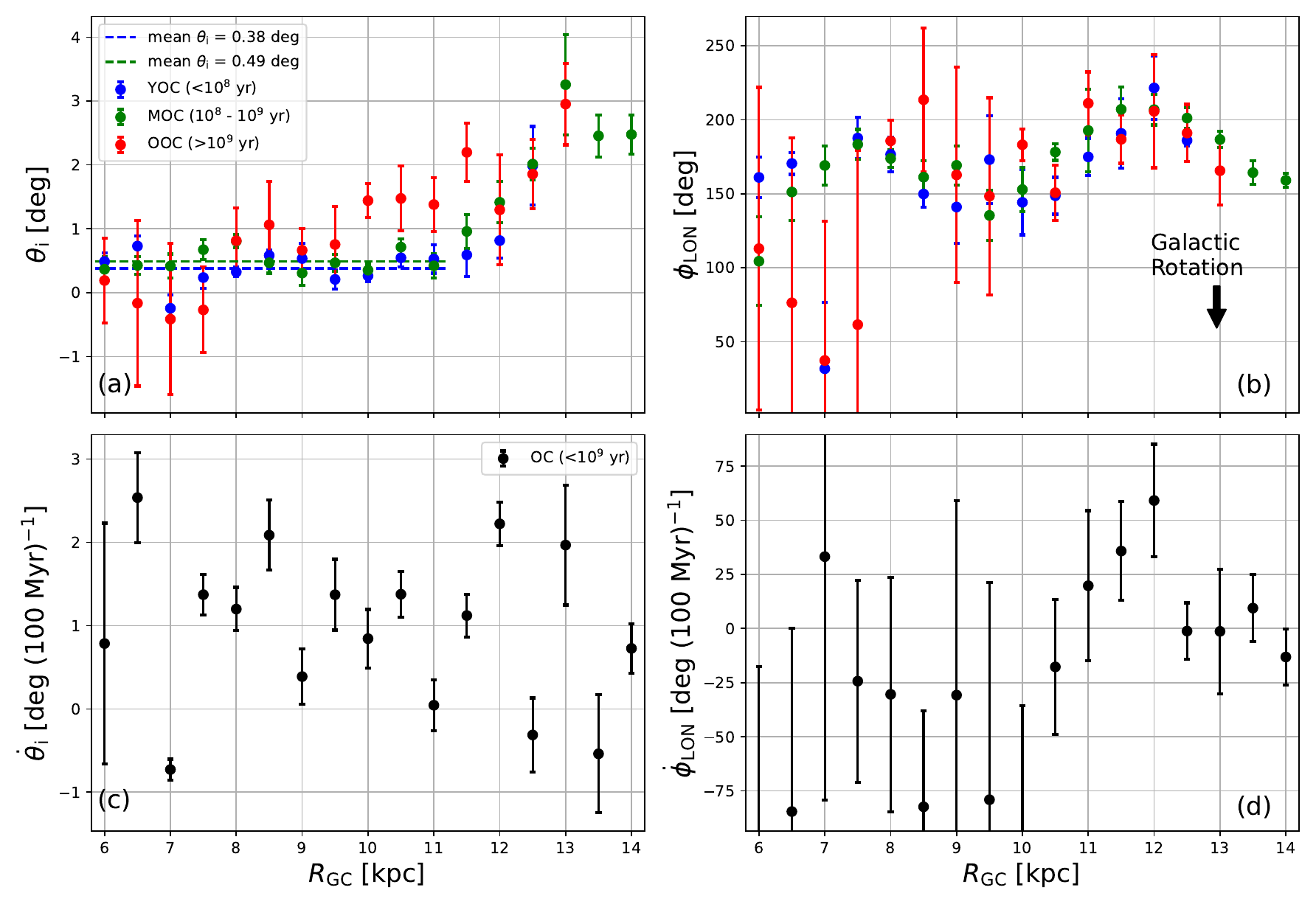}
\caption{
Derived parameters from the dynamical warp of OC samples with different age ranges, $V_{z\odot}$ =~9.43~$\pm$~0.16~\kms is adopted here. The parameters include: (a) inclination angle, (b) LON positions, (c) inclination variation rate, and (d) precession rate, plotted against the Galactocentric radius $\mathrm{R_{GC}}$. The data points with significantly high precession rates ( < -100~deg~yr$^{-1}$) and high uncertainties in the inner disk are outside the displayed range.
}
\label{fig7}
\end{center}
\end{figure*}

Given that tracers of varying ages may exhibit distinct behaviors within warp structures~\citep[e.g.][]{Romero19}.
 Here, we categorize OCs into three age groups to trace the characteristics of warp: young clusters (<~100~Myr, YOCs), middle-aged clusters (0.1 to 1~Gyr, MOCs), and old clusters ($\geq$~1~Gyr, OOCs). Figure~\ref{fig7}-a, we observe that in the middle of the Galactic disk ($\mathrm{R_{GC}}$ = 6 to 11~kpc), there is minimal difference in orbital inclinations between YOCs and MOCs, but in the inner and outer disks, the orbital inclinations of OOCs are greater than those of YOCs. For the most radii, there is a clear trend of increasing inclinations for young and old clusters. Although the differences in orbital inclinations among them fall within the error bars, they exhibit systematic differences across a wide range of Galactocentric distances, indicating a genuine feature. These results are consistent with those of ~\citet{Romero19}, where they found that RGB stars show warping features at smaller Galactocentric distances compared to OB stars.

%%%%%%%%%%%%%%%%%%%%%%%%%%figure%%%%%%%%%%%%%%%%%%%%%%%%%%%%%%%%%
%%%%%%%%%%%%%%%%%%%%%%%%%%figure%%%%%%%%%%%%%%%%%%%%%%%%%%%%%%%%%

Regarding the positions of the LON across three age groups, as illustrated in Figure~\ref{fig7}-b, the fluctuation degree of LON tracked by the OOC surpasses that of both MOC and YOC at various radii. The differences typically range between 10 and 50 degrees. Moreover, the uncertainties associated with OOC are relatively large, likely due to their dispersed distribution and limited sample size. The differences between MOC and YOC are generally minor, with LON discrepancies mostly within 10 degrees.
Interestingly, there appears to be a specific periodicity in the variation of the LON. At a Galactocentric radius of 9.5~kpc, the LON is appoach to the furthest downstream point in the Milky Way's rotation direction. Conversely, at radii of 7.5~kpc and 12~kpc, the LON is situated furthest upstream, and at 14~kpc, it is also approaching the upstream region. This periodicity is more apparent in the distributions of MOC and YOC, although OOC exhibits a comparable trend. If this periodic LON oscillation is confirmed, we speculate based on the observed patterns that the LON might have a variation cycle of approximately 4~kpc along the radius of the Galactic disk.

Meanwhile, based on the $V_{z\odot}$ obtained in this work, we refit the precession rate and inclination variation rate of the warping (lower panels in Figure~\ref{fig7}). In order to minimize uncertainties caused by old OCs, we utilized a star cluster sample with ages less than ~$10^9$ years. The inclination variation rate decreases progressively from the inner disk to the outer disk (Figure~\ref{fig7}-c, also seen in H23), with an approximate change of 1 degree per 100 million years. Over the next 100 to 200 million years, this trend may lead to a reduction in the tilt of both the inner and outer disk. This suggests that the warp of the stellar disk might not be as pronounced in the future as it is currently.
As shown in Figure~\ref{fig7}-d, the data suggests that despite the considerable uncertainty in precession rate, especially notable for $\mathrm{R_{GC}}$ less than 10~kpc due to the low inclination angles, its trend aligns with that of LON. From the most downstream to the most upstream of LON positions, the precession rate gradually decreases and may even become negative. This trend is particularly evident in the outer Galactic disk ($\mathrm{R_{GC}}$ > 10~kpc). Additionally, over time (100 - 200~Myr), the LON values at different $\mathrm{R_{GC}}$ seem to progressively converge towards a shared location.
In addition, the investigations by D23 on Cepheids and H23 on all-aged OCs have both documented a diminishing precession rate of the outer disk warp, with a decrease of 6~\kmsperkpc (35 deg per 100~Myr) between Galactocentric distances of 12~kpc to~14 kpc. Our current results, derived from the solar vertical velocity $V_{z\odot}$, indicate that the precession rates are considerably lower than previously estimated, resulting a reduction of 12.3~\kmsperkpc is observed.

\section{Discussion and Conclusion}\label{sec:5}
%%%%%   1 paragraph to conclusion

In this study, we utilized OCs to investigate warping and found that the tilting of the Galactic disk near the sun leads to vertical systematic motion that affects the magnitude and direction of angular momentum estimates of Galactic objects. By contrasting with the geometric tilt of cluster warping, we obtained the sun's vertical velocity relative to the Galactic mid-plane, $V_{z\odot}$. The results indicate $V_{z\odot}$ = 9.43 $\pm$ 0.16~\kms, which is around 2~\kms larger than the known vertical velocity of the sun ($W_{\odot}$) relative to the local standard of rest. 
The influence of systematic vertical velocities is critical for unraveling the structure and dynamics of the Milky Way, particularly in refining the estimated rate of Galactic precession. Our findings indicate that this rate varies from 10.1~$\pm$~4.4~\kmsperkpc ($\mathrm{R_{GC}}$ = 12~kpc) to -2.2~$\pm$~2.2~\kmsperkpc ($\mathrm{R_{GC}}$ = 14~kpc), with an average values is 1.8~$\pm$~3.3~\kmsperkpc.
These figures are considerably lower than earlier estimates, such as the 10.9~$\pm$~3.2 and 13.6~$\pm$~0.2~\kmsperkpc reported by  ~\citet{Poggio20} and ~\citet{Cheng20}, respectively. 
Such revised rates are essential for more accurate modeling of the Galaxy's dynamical behaviors.

From an observational perspective, warping is a visual phenomenon caused by the average orbital deviation of objects from the mid-plane of the Galactic disk. In fact, our analysis (see Figure~\ref{fig7}-a) reveals that young OCs and some middle-aged OCs at Galactocentric distances of 6 to 11~kpc exhibit very similar inclinations ($\theta_i \sim 0.5^{\circ}$) with respect to the Galactic mid-plane, with their LON close to $\phi$ = 180$^{\circ}$ (i.e., aligned with the line connecting the Galactic center and the sun). This coincidence leads us to consider that the current Galactocentric coordinates may not be the optimal system for studying the warping structure of the Milky Way. If we were to rotate the Galactic mid-plane around the line connecting the Galactic center and the sun by -0.5$^{\circ}$, the inclination of the Galactic middle disk ($\mathrm{R_{GC}}$ = 6 to 11~kpc, including the vicinity of the solar system) traced by young objects may not show any inclinations ($|\theta_\mathrm{i}|$ <~0.1$^{\circ}$ or less). We expected to explore this possibility in future studies.

Through the dynamical orbit distributions of OCs, our study further confirms that older objects on the Galactic disk exhibit larger warping amplitudes compared to younger bodies, both in the outer and inner disk. The inconsistent line of node of OC orbits at different Galactocentric distances reveals the distorted morphology of the disk.  Furthermore, the LONs of the tilted orbit of the Galactic disk exhibits characteristics of periodic oscillations, suggesting that the warping pattern undergoes continuous vibrational changes rather than maintaining a fixed configuration. But the twisted shape of the Galactic disk seems to be recovering, and the warping is also tending to flatten out. 

In our study of the warping structure of the Milky Way, we are still faced with a series of questions. For instance, there is no clear conclusion regarding the origin and dynamical mechanism of the warp, highlighting the urgent need for further investigation to explore potential influencing factors such as the gravitational interactions between the Milky Way and neighboring galaxies, as well as the distribution of dark matter. With the advancement of observational techniques, we anticipate acquiring more precise astrometric data, which will assist in detailing the characteristics of the warping and laying the groundwork for a better understanding of the structure and dynamics of our Galaxy. Simultaneously, delving into the relationship between the warping structure and other features of the Milky Way, such as its interaction with spiral arms, bar, and bulge in the Galactic disk, will provide us with a new perspective on comprehending the overall structure of our Galaxy.

%%%%%%

%===============================================================================================

\begin{acknowledgements}
We appreciate the feedback from Dr. Walter Dehnen, who provided detailed parameter table for CC Warp~\citep{Dehnen23}. This work was supported by National Natural Science Foundation of China through grants 12303024, the Natural Science Foundation of Sichuan Province (2024NSFSC0453), and the “Young Data Scientists” project of the National Astronomical Data Center (NADC2023YDS-07).
This work has made use of data from the European Space Agency (ESA) mission GAIA (\url{https://www.cosmos.esa.int/gaia}), processed by the GAIA Data Processing and Analysis Consortium (DPAC,\url{https://www.cosmos.esa.int/web/gaia/dpac/consortium}). Funding for the DPAC has been provided by national institutions, in particular the institutions participating in the GAIA Multilateral Agreement.

\end{acknowledgements}

\bibliographystyle{aa} % style aa.bst
\bibliography{warp.bib} 
%%%%%%%%%%%%%%%%%%%%%%%%%%%%%%%%%
\begin{appendix}

\section{Sketch map of dynamic warping}\label{sec_sketch_map}
\begin{figure*}[ht]
\begin{center}
\includegraphics[width=0.75\linewidth]{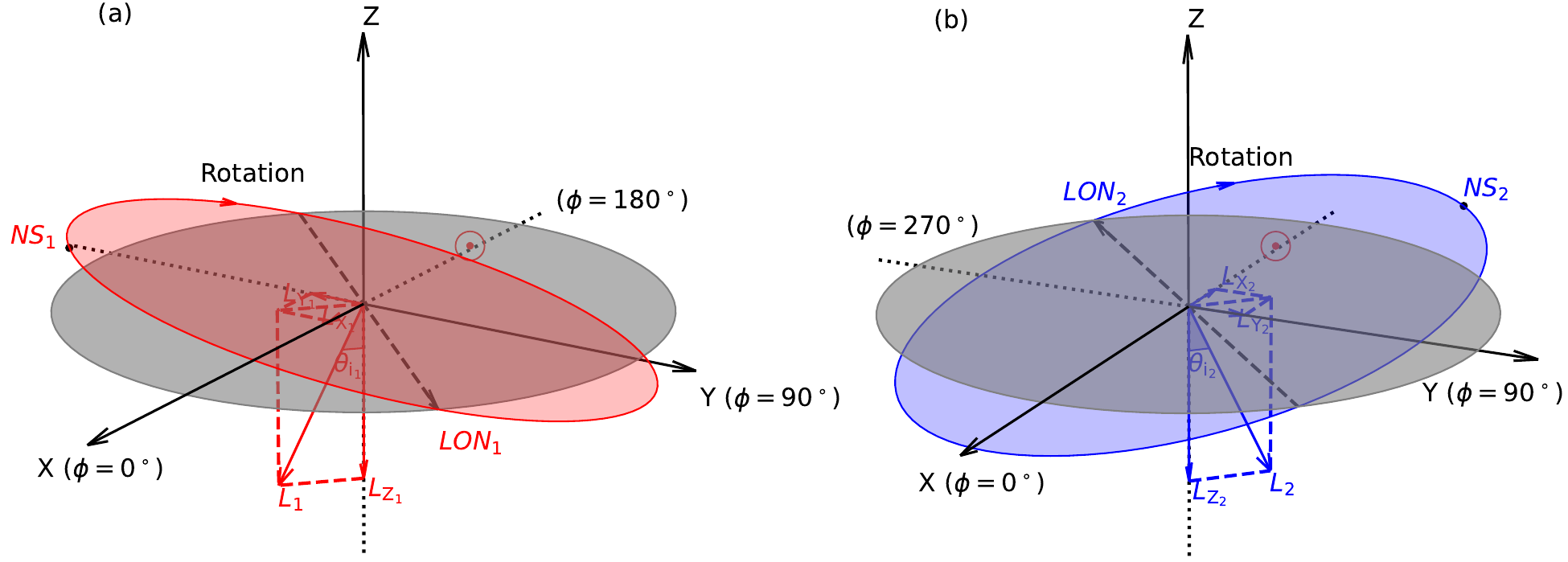}
\caption{Sketch map of the warping disk and OC orbit angular momentum: the grey plane represents the Galactic mid-plane, with the origin located at the Galactic center, and $\phi$ representing the azimuth of the Galactic mid-plane. The blue plane denotes the average instantaneous orbital plane of OCs at a certain range of Galactocentric distance, with the line connecting the intersection of the two planes as the LON (shown by black dashed line, with an arrow indicate the ascending point). $\bm{L}$ represents the normal to the OC orbital plane, while $L_x$, $L_y$, and $L_z$ are the components of $\bm{L}$ in the $X$, $Y$, and $Z$ directions, respectively. The angle $\theta_\mathrm{i}$ between $L_z$ and $\bm{L}$ represents the inclination of the orbital plane. We align the sign of the inclination angle $\theta_\mathrm{i}$ with that of $L_y$, where the negative/positive $L_y$ corresponds to the negative/positive sign of inclination angle, placing the northern solstice (NS, the maximum $Z$) of the warping plane in the third and fourth Galactic quadrants ($l$ and/or $\phi > 180^{\circ}$, $L_y$ < 0, panel a). Conversely, a positive $L_y$ situates the NS in the first and second Galactic quadrants ($l$ and/or $\phi < 180^{\circ}$, panel b). Our coordinate system is the same as that used by H23.}
\label{fig2}
\end{center}
\end{figure*}

We conducted a sketch map of the warping disk traced by  OC orbit angular momentum, as shown in Figure~\ref{fig2}.

\section{Solar peculiar motion values in different works}\label{sec:appendix_a}
We conducted a summary of solar peculiar motion studies over the past two decades, as shown in Table~\ref{table1}.

%%%%%%%%%%%%%%%%%%%%%%%%%%table%%%%%%%%%%%%%%%%%%%%%%%%%%%%%%%%%
\begin{table}[htbp]
\centering
\caption{Published estimations of solar peculiar motions ($U_{\odot}$, $V_{\odot}$, $W_{\odot}$) and rotation speed at the sun's position $\Theta_{0}$. The statistics of peculiar motion before 2019 in the table are from ~\citet{Ding2019}.}
\label{table1}
\begin{tabular}{@{}lcccc@{}}
\toprule
Reference              & $U_{\odot} $ & $V_{\odot}$ & $W_{\odot} $ & $\Theta_0 $ \\
   &[~\kms] & [~\kms] & [~\kms] & [~\kms]\\ \midrule
\cite{Bobylev2007}          & $8.7 \pm 0.5$       & $6.2 \pm 2.2$         & $7.2 \pm 0.8$ &  \\
\citet{Aumer2009}            & $9.96 \pm 0.33$       & $5.25 \pm 0.54$         & $7.07 \pm 0.34$ &  \\
\citet{Reid2009}             & $9$                   & $20$                    & $10$ &  $254 \pm 16$\\
\citet{Francis2009}          & $7.5 \pm 1.00$        & $13.5 \pm 0.30$         & $6.8 \pm 0.10$ &  \\
\citet{Zhu2009}             & $11.7 \pm 0.6$        & $13.3 \pm 0.5$          & $8.1 \pm 0.6$ & $235 \pm 10 $ \\
\citet{Bobylev2010}          & $5.5 \pm 2.2$         & $11 \pm 1.7$            & $8.5 \pm 1.2$ &  $248 \pm 14$\\
\citet{Breddels2010}        & $12 \pm 0.6$          & $20.4 \pm 0.5$          & $7.8 \pm 0.3$ &  \\
\citet{Schonrich2010}       & $11.1^{+0.69}_{-0.75}$& $12.24^{+0.47}_{-0.47}$ & $7.25^{+0.37}_{-0.36}$ &  \\
\citet{Coskunoglu2011}      & $8.5 \pm 0.29$        & $13.38 \pm 0.43$        & $6.49 \pm 0.26$ &  \\
\citet{Golubov2013}          & $8.74 \pm 0.13$       & $3.06 \pm 0.68$         & $7.57 \pm 0.07$ &  \\
\citet{Bobylev2014}          & $6 \pm 0.5$           & $10.6 \pm 0.8$          & $6.5 \pm 0.3$ &  \\
\citet{Reid2014}           & $9.9 \pm 2$       & $14.6 \pm 5$          & $9.3 \pm 1$ & $240 \pm 8$\\
\citet{Sharma2014} 
& $10.96^{+0.14}_{-0.13}$       & $7.53^{+0.16}_{-0.16}$          & $7.539^{+0.095}_{-0.09}$ & $232 \pm 1.7$ \\
\citet{Huang2015}           & $7.01 \pm 0.2$        & $10.13 \pm 0.12$        & $4.95 \pm 0.09$ &  \\
\citet{Tian2015}             & $9.58 \pm 2.39$       & $10.52 \pm 1.96$        & $7.01 \pm 1.67$ &  \\
\citet{Bobylev2016}          & $9.12 \pm 0.1$        & $20.80 \pm 0.1$         & $7.66 \pm 0.08$ & $230 \pm 12$ \\
\citet{Bobylev2017}          & $7.9 \pm 0.65$        & $11.73 \pm 0.77$        & $7.39 \pm 0.62$ &$231 \pm 6 $  \\
\citet{Bobylev17}& $8.19 \pm 0.74$       & $9.28 \pm 0.92$         & $8.79 \pm 0.74$ &$252 \pm 8$  \\
\citet{Ding2019}             & $8.63 \pm 0.64$       & $4.76 \pm 0.49$         & $7.26 \pm 0.36$ &  \\
\citet{Reid19}             & $10.6 \pm 1.2$        & $10.7 \pm 6.0$             & $7.6 \pm 0.7$ &  $236 \pm 7$\\
\citet{Bobylev2019}         & $8.53 \pm 0.38$       & $11.22 \pm 0.46$        & $7.83 \pm 0.32$ &  $229.7 \pm 4.6$  \\
\citet{Bobylev2019AstL}         & $7.88 \pm 0.48$       & $11.17 \pm 0.63$        & $8.28 \pm 0.45$ & $235 \pm 5$ \\
\citet{Wang2021}           & $11.69 \pm 0.68$      & $10.16 \pm 0.51$        & $7.67 \pm 0.10$ & $231.47 \pm 9.81$ \\
\citet{Kobulnicky2022}            & $5.50 \pm 1.00$       & $7.50 \pm 1.00$         & $4.50 \pm 1.00$ &  \\
\citet{Guo2023}         & $10.10 \pm 0.10$      & $22.80 \pm 0.10$        & $7.80 \pm 0.10$ & $251.2 \pm 19$ \\
\citet{Semczuk23}         & 13     &     & 6.9    & \\   
\citet{gc20210};~\citet{Drimmel23} & $9.3 \pm 1.3$ & & $8.59 \pm 0.28$ & \\
\bottomrule
\end{tabular}
\end{table}

\end{appendix}

\end{document}